\documentclass[a4paper,american,british,english,3p,times,twocolumn,floatfix]{revtex4-1}
\usepackage[T1]{fontenc}
\usepackage[latin9]{inputenc}
\usepackage{units}
\usepackage{graphicx}
\usepackage{esint}

\makeatletter

\usepackage{hyperref}
\DeclareMathAlphabet\mathbfcal{OMS}{cmsy}{b}{n}

\makeatother

\usepackage{babel}
\begin{document}

\title{Propagating Cosmic Rays with exact Solution of Fokker-Planck Equation}
\selectlanguage{british}%

\author{M.A. Malkov}

\address{University of California, San Diego, La Jolla, CA 92093}
\selectlanguage{english}%
\begin{abstract}
Shortfalls in cosmic ray (CR) propagation models obscure the CR sources
and acceleration mechanisms. This problem became particularly obvious
after the Fermi, Pamela, and AMS-02 have discovered the electron/positron
and $p/$He spectral anomalies. Most of the CR models use diffusive
propagation that is inaccurate for weakly scattered energetic particles.
So, some parts of the spectra affected by the heliospheric modulation,
for example, cannot be interpreted. I discuss and adopt an exact solution
of the Fokker-Planck equation \cite{2016arXiv161001584M}, which gives
a complete description of a ballistic, diffusive and transdiffusive
(intermediate between the first two) propagation regimes. I derive
a simplified version of an exact Fokker-Planck propagator that can
easily be employed in place of the Gaussian propagator, currently
used in major Solar modulation and other CR transport models.
\end{abstract}

\maketitle

\section{Lacuna in CR Transport Model}

The cosmic ray (CR) propagation in turbulent environments, such as
the interstellar medium (ISM) or Heliosphere, has been actively researched
for more than half a century \cite{Jok1971RvGSP}. Time asymptotically,
CRs propagate diffusively; after several collisions, they \textquotedblleft forget\textquotedblright{}
their initial velocities and enter a random walk process. However,
in astrophysical objects with infrequent particle collisions, there
may not be enough time or room for even a few collisions. In such
systems, the focus shifts to earlier propagation phases, which are
better described as ballistic rather than diffusive propagation. The
question is, what is in between these two regimes and how long it
lasts?

The transition from ballistic to diffusive transport regime has always
been a challenge for the theory. At the same time, it is often the
key to understanding the CR sources. Since the particle mean free
path (m.f.p) usually grows with energy, some part of their spectrum
almost inescapably falls into a transient category where neither ballistic
nor diffusive approximation applies. I will call this regime transdiffusive
and argue that it lasts for long enough to compromise both the ballistic
and diffusive model predictions. During this propagation phase, CR
protons accelerated in supernova remnants (SNR), for example, may
reach a nearby molecular cloud, making themselves visible by interacting
with its dense gas \cite{MDS_11NatCo,AgileW44_11}. The CR protons
of lower energies would instead be diffusively confined to the SNR
shell and evade detection. Due to a high CR intensity near the source,
however, their confinement here must be due to self-generated Alfven
waves. At a minimum, this problem should be treated at a quasilinear
level \cite{MetalEsc13}, as opposed to the linear CR transport, considered
throughout this paper. Another example is the propagation of solar
energetic particles to 1 AU. Also, in this case, the m.f.p. of some
particles is comparable to, or even exceed, 1 AU, so neither the diffusive
nor ballistic approximation applies \cite{Klassen2016A&A,Bian2017ApJ}.

Galactic CRs ultimately propagating through the Heliosphere to the
observer cannot always be propagated back to their source within simple
diffusion or ballistic paradigms, so their spectra cannot be fully
understood. This problem is particularly relevant to striking\foreignlanguage{british}{
anomalies in the CR spectra and composition, which are becoming a
general trend in the CR observations. Besides the $e^{+}/e^{-}$ anomaly,
there is a $\sim0.1$ difference in rigidity indices of proton and
He. Although the explanations are available (see, e.g., \cite{MDSPamela12,MalkPositrons2016},
and a companion paper in this volume), the low-energy parts of these
spectra are strongly affected by the solar modulation. Curvature and
gradient drifts in the Heliospheric magnetic field are mostly treated
by considering particle propagation along the field line as diffusive,
e.g. \cite{Potgieter2013}, which we will show to be inaccurate for
sufficiently energetic particles with long m.f.p.}

\section{Governing Equation}

The Fokker-Planck (FP) equation is a minimalist model suitable for
the CR transport. An ambient magnetic field justifies a 1D treatment,
while its fluctuating part supports the particle scattering in pitch
angle. The simplest form of FP equation for the CR distribution function
$f$ is the following:

\begin{equation}
\frac{\partial f}{\partial t}+v\mu\frac{\partial f}{\partial x}=\frac{\partial}{\partial\mu}\left(1-\mu^{2}\right)D\left(\mu,E\right)\frac{\partial f}{\partial\mu}.\label{eq:PAscatIntro}
\end{equation}
Here $x$ is directed along the local magnetic field, $\mu$ is the
cosine of the particle pitch angle, $v,E$ are the particle velocity
and energy, conserved in interactions with quasi-static magnetic turbulence.
$D$ is the scattering rate (collision frequency). 

One propagation scenario that Eq.(\ref{eq:PAscatIntro}) describes
very well comes about through an instant release of a cloud of particles
into a scattering medium. Again, Galactic SNRs, widely believed to
generate CRs with energies up to $\sim10^{15}$eV, must accelerate
them in SNR shock waves with a subsequent release into a turbulent
ISM. The question then is how exactly the particle density (the isotropic
component of $f$) propagates along a magnetic flux tube that intersects
the SNR shell. The goal is to achieve the simplicity of diffusive
description (e.g., \cite{Jokipii66} and below) which is a well-known
derivative of Eq.(\ref{eq:PAscatIntro}). As emphasized earlier, the
diffusive treatment is inadequate in the preceding ballistic and transdiffusive
propagation phases, while the latter is often the key for probing
into the source. 

\subsection{Restricting Propagation Models by Limiting Cases \label{subsec:Restricting-Propagation-Models}}

Because of the difficulties in reducing the FP equation to a manageable
isotropic form, a framework for such reduction limited by the extreme
cases of ballistic and diffusive propagation is helpful. We derive
both regimes directly from Eq.(\ref{eq:PAscatIntro}), by eliminating
angular dynamics. 

In the ballistic case, which strictly applies to times shorter than
the collision time  $t\ll t_{c}\sim1/D$, one can neglect the r.h.s.
altogether. The solution then follows from integrating along the particle
trajectories, $x-\mu vt=const$ (Liouville's theorem), with a conserved
pitch angle, $\mu=const$. The solution is simply $f\left(x,\mu,t\right)=f\left(x-v\mu t,\mu,0\right)$. 

Consider an isotropic point source: $f\left(x,\mu,0\right)=\nicefrac{1}{2}\delta\left(x\right)\Theta\left(1-\mu^{2}\right),$
where $\delta$ and $\Theta$ denote the Dirac's delta and Heaviside
unit step functions, respectively. From the above solution for $f\left(x,\mu,t\right)$,
one obtains the ballistic expansion in form of the second moment,
$\left\langle x^{2}\right\rangle =v^{2}t^{2}/3$ by integrating $x^{2}f=\nicefrac{1}{2}x^{2}\delta\left(x-v\mu t\right)\Theta\left(1-\mu^{2}\right)$
over $x$ and $\mu$. The result describes a free escape with the
mean square velocity $v/\sqrt{3}$, while the maximum particle velocity
(along $x$) is $v$. The pitch angle averaged particle distribution,
$f_{0}\left(x,t\right)=\left(2vt\right)^{-1}\Theta\left(1-x^{2}/v^{2}t^{2}\right)$,
is best described as an expanding 'box' of decreasing height. Among
earlier attempts to reduce $f$ to its pitch angle-averaged part,
$f\left(t,x,\mu\right)\to f_{0}\left(t,x\right)$, an approach leading
to a ``telegraph'' equation, can be readily tested using the above
box solution. We will briefly discuss this approach later and show
that it is inconsistent with the ballistic limit of $f_{0}$ obtained
directly from the FP equation. Needless to say that the exact solution
of Eq.(\ref{eq:PAscatIntro}), presented further in this paper, converges
to the above-described box distribution at $t\ll t_{c}$.

The second, well studied propagation regime is diffusive. It dominates
at $t\gg t_{c}\sim1/D$ and is treated in a way opposite to the above-described
ballistic regime, \cite{Jok1971RvGSP}. The r.h.s. of Eq.(\ref{eq:PAscatIntro})
is now the leading term, thus implying that the particle distribution
is close to isotropy, $\partial f/\partial\mu\to0$. Working to higher
orders in anisotropic corrections $\sim1/D$, and averaging the equation
over $\mu$, one obtains the following equation for $f_{0}\left(x,t\right)$
\cite{MS2015}

\begin{equation}
\frac{\partial f_{0}}{\partial t}-\kappa_{2}\frac{\partial^{2}f_{0}}{\partial x^{2}}=-\kappa_{4}\frac{\partial^{4}f_{0}}{\partial x^{4}}+\kappa_{6}\frac{\partial^{6}f_{0}}{\partial x^{6}}-\dots,\label{eq:AppendDiffHyperdiff}
\end{equation}
with $\kappa_{2n}\sim1/D^{n}$. The last equation results from an
asymptotic (Chapman-Enskog) expansion of the problem in $1/D$ under
the scattering symmetry: $D\left(-\mu\right)=D\left(\mu\right)$.
It is valid only for $t\gg t_{c}\sim1/D$, and all the short-time-scale,
ballistic propagation effects are intentionally eliminated (cf. elimination
of secular terms in perturbative treatments). A failure to do so results
in a second order time derivative in Eq.(\ref{eq:AppendDiffHyperdiff})
(already mentioned telegraph term) which is illegitimate unless $t\gg t_{c}$.
Nevertheless, the telegraph equation has been putting forward over
the last 50 years as a viable tool for describing the CR propagation
from the ballistic to diffusive phases.

\begin{figure}
\includegraphics[bb=0bp 0bp 650bp 480, scale=0.6]{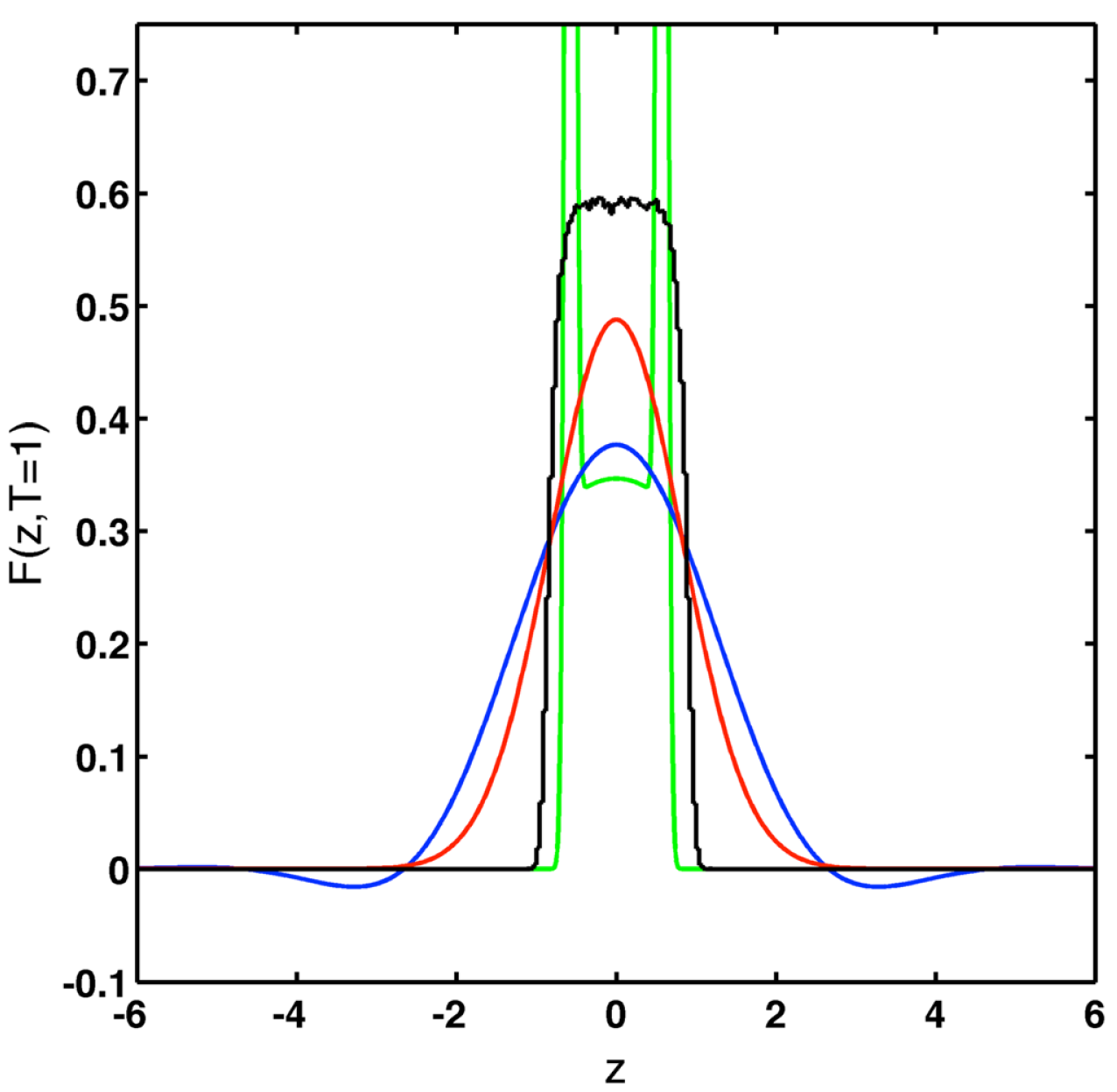}

\caption{Adopted from Ref.\cite{Litvin2016PhPl} are the solutions of the FP
equation (black line), telegraph (green), diffusion/hyperdiffusion
(red/blue). The latter two are not applicable for this short ($t=t_{c}$)
integration time as they converge to the FP solution very slowly (see
fig.\ref{fig:SixPlots} below and \cite{2016arXiv161001584M}).\label{fig:Telegraph} }
\end{figure}

Meanwhile, the r.h.s. of eq.(\ref{eq:AppendDiffHyperdiff}) provides
small hyperdiffusive corrections that may be omitted at $t>t_{c}$,
as the higher spatial derivatives quickly decay because of the smoothing
effect from the diffusive term on its l.h.s. These corrections do
not shed much light on the ballistic and transdiffusive propagation
regimes, probably unless the series is summed up with no truncation.
The latter requirement derives from a method whereby an exact solution
of the parent FP equation \cite{2016arXiv161001584M} is evaluated.
The evaluation consists in summing up an \emph{infinite} series of
moments $\left\langle x^{2n}f_{0}\left(x,t\right)\right\rangle $
that are evidently connected with the infinite series of coefficients
$\left\{ \kappa_{2n}\right\} $ in eq.(\ref{eq:AppendDiffHyperdiff}).
Conversely, by including just one (or several) hyperdiffusive correction
outside of their validity range, $t\gg t_{c}$, one may even decrease
the accuracy of the diffusive approximation. It can also be shown
\cite{MS2015} that within its validity range, a truncated version
of eq.(\ref{eq:AppendDiffHyperdiff}), with $\kappa_{2n}=0$ for $n>2$,
can be mapped onto the telegraph equation. It follows that neither
a \emph{truncated} hyperdiffusive approach nor the telegraph equation
(a subset of the former) cannot adequately reproduce the FP solution
at times shorter than $t\gg t_{c}$. This was recently demonstrated
in Ref.\foreignlanguage{american}{\cite{Litvin2016PhPl}}, by a numerical
integration of Eq.(\ref{eq:PAscatIntro}). The results of this work
are illustrated for $t=t_{c}$ in Fig.\ref{fig:Telegraph}. We will
quantify the constraint $t\gg t_{c}$, repeatedly stressed above,
by comparing the full FP solution with its diffusive limit (see \cite{2016arXiv161001584M}
for more details).

The primary failure of the diffusive approach is an unrealistically
fast (acausal) propagation, which is especially pronounced during
the ballistic and transdiffusive phases. Mathematically, the approximation
violates an upper bound $\left|x\right|\le vt$ that immediately follows
from Eq.(\ref{eq:PAscatIntro}) for a point source solution, discussed
above. There have been attempts to overcome this problem, but no adequate
\emph{ab initio} description of particle spreading that would cover
ballistic and diffusive phases was elaborated. The most persistent
such attempt is based on the telegraph equation discussed above. It
has a misleading impact on the field of CR propagation for that simple
reason that the solution of this equation is inconsistent with its
parent FP equation. We obtained this simple result by considering
the ballistic propagation phase directly from eq.(\ref{eq:PAscatIntro})
(see \cite{MS2015,M_PoP2015,2016arXiv161001584M} for more discussion).

It follows that there are no viable analytical tools to address the
earlier phases of particle propagation, except to possibly sum up
the series of hyperdiffusive terms or just to solve the FP equation
directly. Below, we take the second option. 

\section{Exact Solution of FP equation}

The energy dependence of the particle scattering frequency enters
eq.(\ref{eq:PAscatIntro}) only as a \emph{parameter}, i.e., $D\left(E\right)$.
The possible pitch-angle dependence of $D$ typically scales as $D\left(\mu\right)\propto\left|\mu\right|^{q-1}$
\cite{Jok1971RvGSP}, thus being suppressed in an important case $q=1$,
where $q$ is the power-law index of magnetic turbulence. Under these,
quite realistic assumptions, the FP equation can be solved exactly
\cite{2016arXiv161001584M}. To describe this solution, it is convenient
to rewrite Eq.(\ref{eq:PAscatIntro}) using dimensionless time and
length units according to the following transformations

\begin{equation}
D\left(E\right)t\to t,\;\;\;\frac{D}{v}x\to x\label{eq:RescaledTimeANDx}
\end{equation}
Instead of Eq.(\ref{eq:PAscatIntro}) we thus have

\begin{equation}
\frac{\partial f}{\partial t}+\mu\frac{\partial f}{\partial x}=\frac{\partial}{\partial\mu}\left(1-\mu^{2}\right)\frac{\partial f}{\partial\mu}\label{eq:FPundim}
\end{equation}
This equation contains no parameters, thus precluding any direct asymptotic
expansion in a small parameter, unless it enters the problem implicitly
through the initial condition $f\left(x,\mu,0\right)$. In particular,
if one is using Eq.(\ref{eq:AppendDiffHyperdiff}) ($1/D$- type expansion),
not only should the initial distribution be close to isotropy, but
it should also be spatially broad. The latter condition will prevent
a high anisotropy from arising via the second term on the l.h.s. of
Eq.(\ref{eq:FPundim}). Hence, the problem of a point source spreading
(Green's function, or fundamental solution) can not be treated using
conventional $1/D$ expansion, until $f$ becomes quasi-isotropic,
that is broadened to $x\gtrsim1$.

The exact solution of Eq.(\ref{eq:FPundim}) can be obtained using
a fully resolvable infinite set of moments of $f\left(\mu,x\right)$

\begin{equation}
M_{ij}\left(t\right)=\left\langle \mu^{i}x^{j}\right\rangle =\int_{-\infty}^{\infty}dx\int_{-1}^{1}\mu^{i}x^{j}fd\mu/2\label{eq:MomentsDef}
\end{equation}
for any integer $i,j\ge0$. The lowest moment $M_{00}$ is automatically
conserved by Eq.(\ref{eq:FPundim}) (as being proportional to the
number of particles) and we normalize it to unity, $M_{00}=1$. All
the higher moments can be explicitly obtained from the following recurrence
relation 

\[
M_{ij}\left(t\right)=M_{ij}\left(0\right)e^{-i\left(i+1\right)t}+\int_{0}^{t}e^{i\left(i+1\right)\left(t^{\prime}-t\right)}\times
\]

\begin{equation}
\left[jM_{i+1,j-1}\left(t^{\prime}\right)+i\left(i-1\right)M_{i-2,j}\left(t^{\prime}\right)\right]dt^{\prime}\label{eq:MomSol}
\end{equation}
Focusing on a point source (fundamental) solution, we assume the initial
distribution $f\left(x,\mu,0\right)$ to be symmetric in $x$ and
isotropic in $\mu$ which eliminates the odd moments. Furthermore,
the initial spatial width must then also be set to zero, $M_{02}\left(0\right)=\left\langle x^{2}\right\rangle _{0}=0$. 

From the mathematical point of view, only a \emph{full} set (first two moments have been calculated by G.I. Taylor \cite{Taylor01011922})
of moments in Eq.(\ref{eq:MomSol}) provides a complete solution $f\left(x,\mu,t\right)$
of Eq.(\ref{eq:FPundim}) given the initial value, $f\left(x,\mu,0\right)$
that determines the matrix $M_{ij}\left(0\right)$ in Eq.(\ref{eq:MomSol}).
Moreover, to adequately reproduce the ballistic and transdiffusive
phases the series of moments cannot be truncated. Considering the
fundamental solution, we will focus on the isotropic part of particle
distribution 

\begin{equation}
f_{0}\left(x,t\right)=\int_{-1}^{1}f\left(\mu,x,t\right)d\mu/2,\label{eq:f0Def}
\end{equation}
as only this part contributes to the particle number density. To obtain
the fundamental solution we impose the initial condition $f_{0}\left(x,0\right)=\delta\left(x\right)$.
The matrix elements that represent $f_{0}$ are, therefore, $M_{0,j}$,
which we denote $M_{j}$: 

\[
M_{j}\equiv M_{0,j}
\]
Note, that $M_{ij}$ with $i>0$ are not small and remain essential
for calculating the full set of the moments $M_{j}$. To link them
to $f_{0}$, we use the moment-generating function 

\begin{equation}
f_{\lambda}\left(t\right)=\int_{-\infty}^{\infty}f_{0}\left(x,t\right)e^{\lambda x}dx=\sum_{n=0}^{\infty}\frac{\lambda^{2n}}{\left(2n\right)!}M_{2n}\left(t\right)\label{eq:MomGenFunc}
\end{equation}
where we omitted the odd moments irrelevant to the fundamental (symmetric
in $x$) solution. The above expansion may be cast in a familiar Fourier
transform of $f_{0}$$\left(x,t\right)$ by setting $\lambda=-ik$. 

Since expressions for the moments $M_{2n}$ are becoming cumbersome
with growing $n$, an exact form the Green's function $f_{0}\left(x,t\right)$,
which can be recovered from eq.(\ref{eq:MomGenFunc}) by inverting
the Fourier integral, is also not simple. Therefore, in the next section,
we derive a new simplified version of the exact FP propagator that
was recently obtained in Ref. \cite{2016arXiv161001584M}.

\selectlanguage{american}%
\begin{figure}
\selectlanguage{english}%
\includegraphics[bb=0bp 120bp 650bp 700bp, scale=0.37]{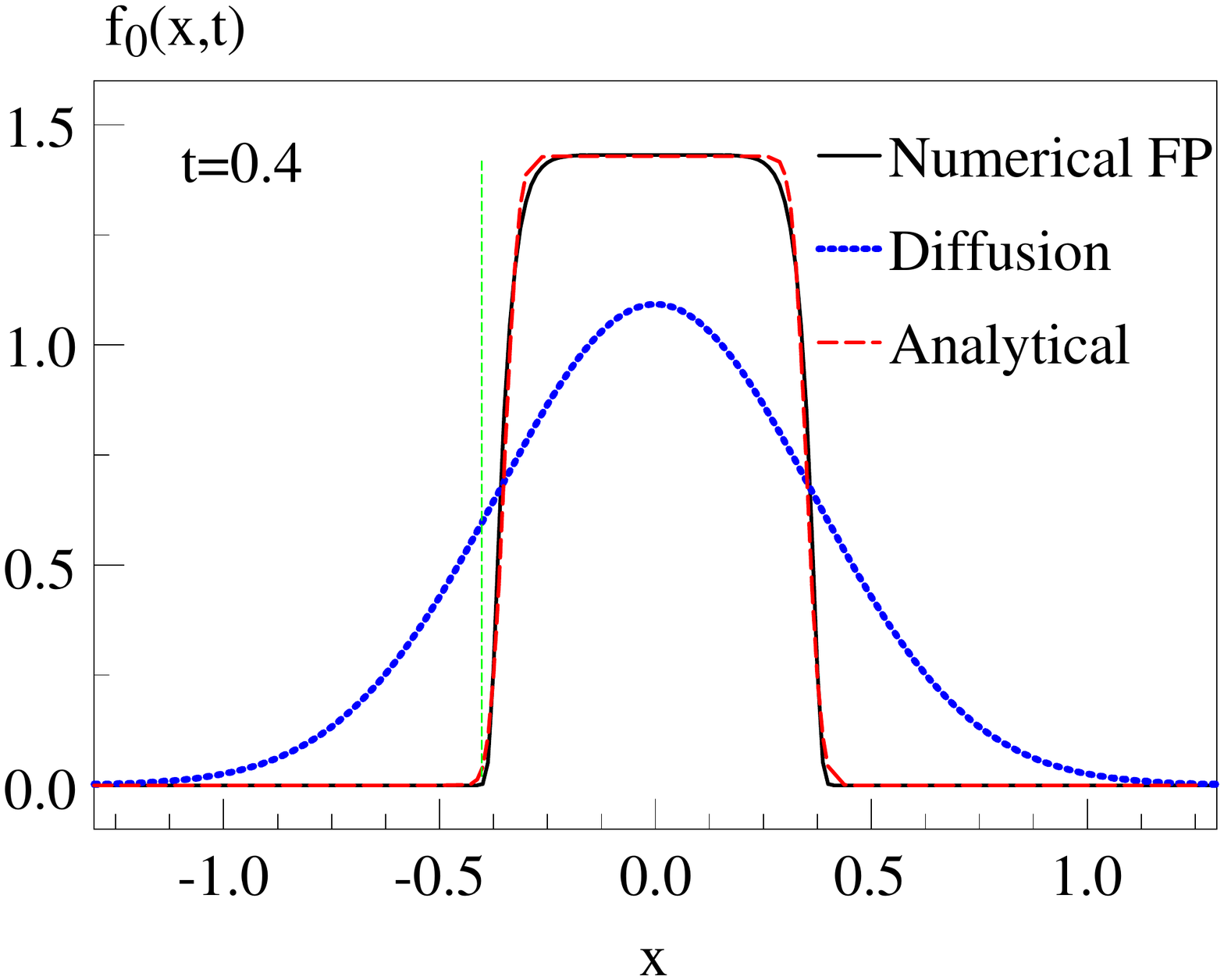}
\includegraphics[bb=0bp  120bp 650bp 680bp, scale=0.37]{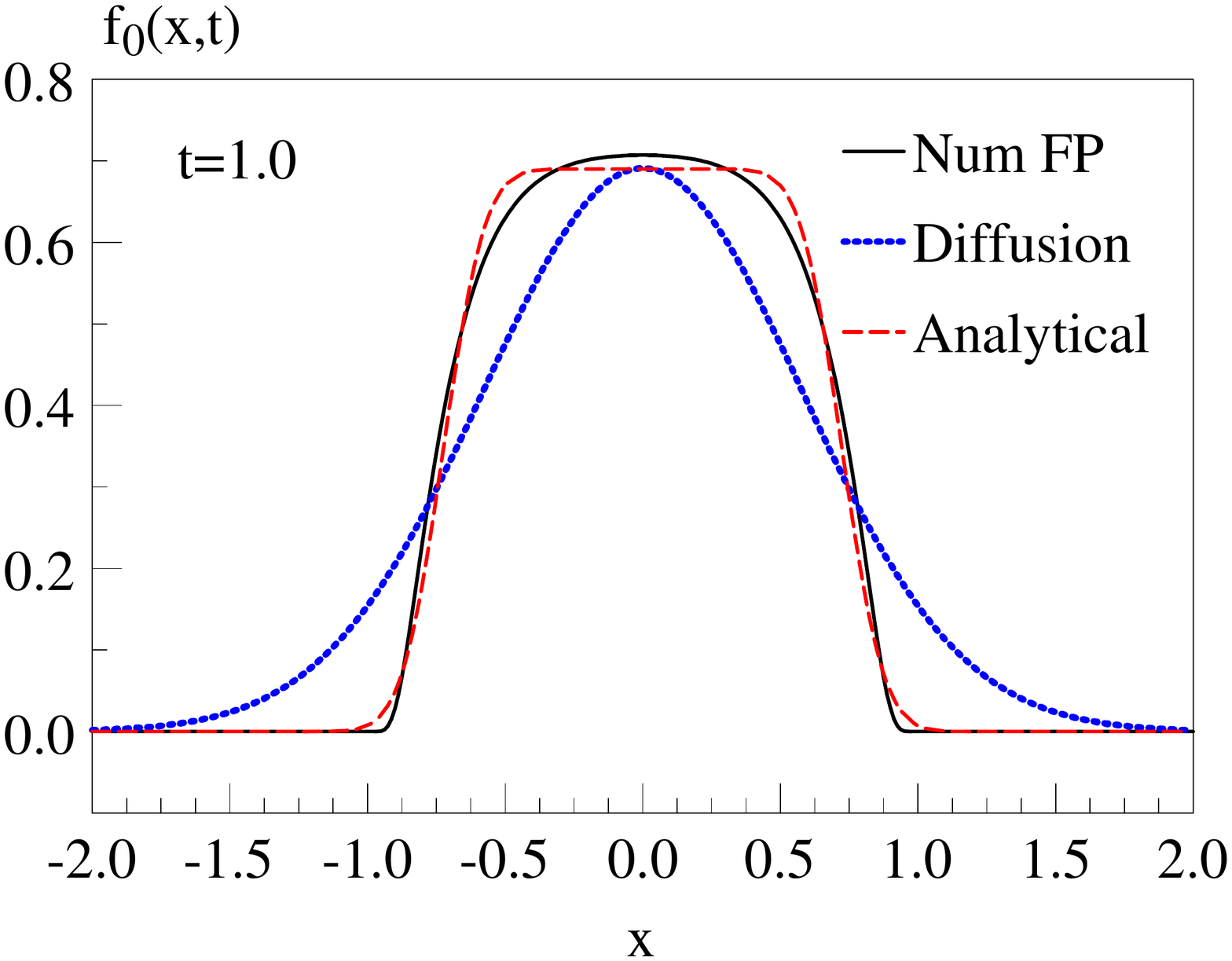}
\includegraphics[bb=0bp 80bp 650bp 680bp,scale=0.37]{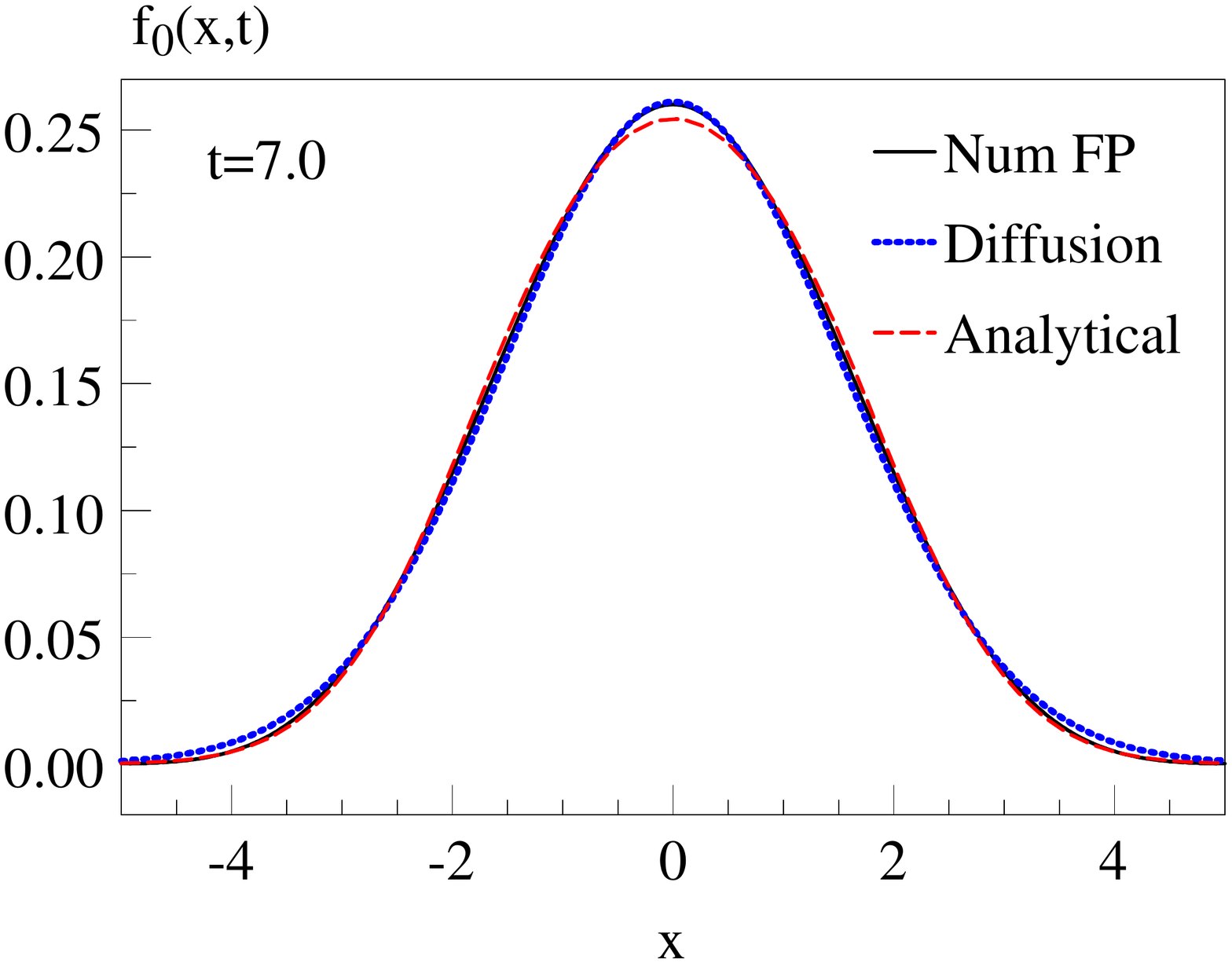}

\selectlanguage{american}%
\caption{Fundamental solution of the Fokker-Planck equation shown for its isotropic
component, $f_{0}\left(x,t\right)=\left\langle f\left(x,\mu,t\right)\right\rangle $
at $t=0.4,\;1.0,\;7.0$. Analytic approximation is from Eq.(\ref{eq:f0Univers}),
diffusive (Gaussian) solution from Eq.(\ref{eq:Gauss}), numerical
- from the FP eq.(\ref{eq:FPundim}). Vertical green line in the upper
panel shows the width of the front. \label{fig:SixPlots}}
\end{figure}

\selectlanguage{english}%

\section{Two-moment Fokker-Plank Propagator\label{sec:Two-moment-FP-Propagator}}

The infinite series entering the moment generating function $f_{\lambda}$
in eq.(\ref{eq:MomGenFunc}) has been summed up by considering the
cases of small and large values of $t$ and $\lambda t$. Despite
the multiplicity of limiting cases associated with these two independent
quantities, all the expressions for the sum $f_{\lambda}\left(t\right)$
are surprisingly similar. They can be unified under a single \emph{approximate
}(but\emph{ valid }for all $x$ and $t$) expression for $f_{0}$:

\begin{equation}
f_{0}\left(x,t\right)\approx\frac{1}{4y}\left[{\rm erf}\left(\frac{x+y}{\Delta}\right)-{\rm erf}\left(\frac{x-y}{\Delta}\right)\right].\label{eq:f0Univers}
\end{equation}
It has been obtained in \cite{2016arXiv161001584M} from an inverse
Fourier transform, $f_{\lambda}\left(t\right)\mapsto f_{0}\left(x,t\right)$,
after summing up the series for $f_{\lambda}$ in eq.(\ref{eq:MomGenFunc}).
The two independent functions of time, $y\left(t\right)$ and $\Delta\left(t\right)$
can be expressed through the moment $M_{2}\left(t\right)$, which
we calculate exactly from eq.(\ref{eq:MomSol}). The solution $f_{0}$
with $y$ and $\Delta$ so obtained compares very well with the numerical
FP solution. The disadvantage of this single-moment representation
of $y$ and $\Delta$ is that it requires some (fairly minor, though)
changes in $y\left(M_{2}\right)$ and $\Delta\left(M_{2}\right)$,
between the cases $t\lesssim1$ and $t\gtrsim1$ \cite{2016arXiv161001584M}.

Here we suggest an alternative representation of the functions $y$
and $\Delta$. Although they lead to a slightly less accurate value
of $f_{0}$ at $t\sim1$ in eq.(\ref{eq:f0Univers}), but are the
same for arbitrary $t:$ $0<t<\infty$. The idea behind this method
of determination of $y$ and $\Delta$ is very simple. As the general
form of the solution given in eq.(\ref{eq:f0Univers}) must arguably
be the same for all $t$, we find the functions $y\left(t\right)$
and $\Delta\left(t\right)$ by requiring that $f_{0}\left(x,t\right)$
exactly satisfies the following two relations

\begin{equation}
M_{2}=\int x^{2}f_{0}\left(x,t\right)dx,\;\;M_{4}=\int x^{4}f_{0}\left(x,t\right)dx\;\;\label{eq:M2M4}
\end{equation}
Recall, that we know exact values for all moments $M_{n}$ from eq.(\ref{eq:MomSol}).
Here, we will only use $M_{2}$ and $M_{4}$, which satisfy the initial
conditions, $M_{2}\left(0\right)=0$ and $M_{4}\left(0\right)=0$: 

\[
M_{2}=\frac{t}{3}-\frac{1}{6}\left(1-e^{-2t}\right)
\]

\[
M_{4}=\frac{1}{270}e^{-6t}-\frac{t+2}{5}e^{-2t}+\frac{1}{3}t^{2}-\frac{26}{45}t+\frac{107}{270}
\]
Substituting $f_{0}$ from eq.( \ref{eq:f0Univers}) into eqs.(\ref{eq:M2M4}),
we find

\begin{equation}
y=\left[\frac{45}{2}\left(M_{2}^{2}-\frac{1}{3}M_{4}\right)\right]^{1/4}\label{eq:y}
\end{equation}

\begin{equation}
\Delta=\sqrt{2M_{2}-\sqrt{10}\sqrt{M_{2}^{2}-\frac{1}{3}M_{4}}}\label{eq:Delta}
\end{equation}

The FP solution, cast in a simplified form of eq.(\ref{eq:f0Univers}),
is not more difficult than the familiar diffusive solution. If we
ignore, for a moment, the time dependence of $y$ in the error functions,
the FP solution appears as the solution of a conventional diffusion
problem with an initial particle density evenly distributed between
$-y<x<y$, and zero otherwise. The essential difference is only in
the form of $y\left(t\right)$ and $\Delta\left(t\right)$. The first
notable aspect of this solution is that at $t\ll1$ it exactly corresponds
to an 'expanding box' ballistic solution described in Sec. \ref{subsec:Restricting-Propagation-Models}.
Indeed, since $\Delta\propto t^{2}$ and $y\approx t$ for $t\ll1$,
the difference of the two error functions yields $2\Theta\left(1-x^{2}/t^{2}\right)$
and $f_{0}$ in eq.(\ref{eq:f0Univers}) is the same as the expanding
box solution obtained in Sec.\ref{subsec:Restricting-Propagation-Models}.
The telegraph solution, on the contrary, is inconsistent with this
expansion regime as it contains two (nonexistent in the FP solution)
singular components at the two propagating fronts, let alone the front
positions and the overall profile, Fig\ref{fig:Telegraph}.

The width of the propagating fronts at $x=\pm y$, determined by $\Delta\left(t\right)$,
behaves as follows, Fig.\ref{fig:SixPlots}. At small $t\ll1$, when
the box is expanding ballistically, i.e. $y\approx t$, the wall thickness
$\Delta\approx2t^{2}/3\sqrt{5}.$ After gradually proceeding through
the transdiffusive phase, these quantities become $y\approx\left(11t/6\right)^{1/4}$
and $\Delta\approx\left(2t/3\right)^{1/2}$ for $t\gg1$. Accordingly,
the expression in eq.(\ref{eq:f0Univers}) converges (rather slowly,
though) to:

\selectlanguage{american}%
\begin{equation}
f_{0}\left(x,t\right)=\sqrt{\frac{3}{2\pi t}}e^{-3x^{2}/2t}\label{eq:Gauss}
\end{equation}
which is the diffusive asymptotic solution of the pitch angle averaged
FP equation, given by eq.(\ref{eq:AppendDiffHyperdiff}) with $\kappa_{2}=1/6$
and all the hyperdiffusive coefficients $\kappa_{2n}=0$ for $n>1$.
Summarizing this section, the two-moment single formula representation
of the FP solution in eq.(\ref{eq:f0Univers}) has correct asymptotic
limits at $t\to0,\infty$, both obtained independently. The remaining
deviations from the numerical solution at $t\sim1$ are minor and
more than compensated by the simplicity of eq.(\ref{eq:f0Univers})
and its validity for all $0<t<\infty$.

\section{Conclusions}

\selectlanguage{english}%
The exact solution of FP equation obtained in \cite{2016arXiv161001584M}
is transformed into a simple form that accurately evolves the pitch
angle averaged particle distribution $f_{0}\left(x,t\right)$, uniformly
in $-\infty<x<\infty$ and $0\le t<\infty$. 

The overall CR propagation can be categorized into three phases: ballistic
($t<1)$, transdiffusive ($t\sim1$) and diffusive ($t\gg1$), (time
in units of collision time $t_{c}$). In the ballistic phase, the
source expands as a ``box'' of size $\Delta x\propto\sqrt{\left\langle x^{2}\right\rangle }\propto t$
with thickening ``walls'' at $x=\pm y\left(t\right)\approx\pm t$
of the width $\Delta\propto t^{2}$. The next, transdiffusive phase
is marked by the box's walls thickened to a sizable fraction of the
box $\Delta\sim\Delta x\sim y$ and its slower expansion, Fig. \ref{fig:SixPlots}.
Finally, the evolution enters the conventional diffusion phase, in
which $\Delta x\sim\Delta\propto\sqrt{t}$, while the walls are completely
smeared out, as $y\propto t^{1/4}$, so $y\ll\Delta$. 

In constraining earlier FP-based models for the CR propagation, the
exact FP solution reveals the following:
\begin{itemize}
\item the conventional diffusion approximation can be safely applied but,
only after 5-7 collision times, depending on the accuracy requirements
\item a popular telegraph approach, originally intended to cover also the
earlier propagation phases at $t\lesssim1$, is inconsistent with
the exact FP solution (see also \cite{2016arXiv161001584M})
\item no signatures of (sub) super-diffusive propagation regimes are present
in the exact FP solution 
\end{itemize}
The latter regimes are occasionally postulated, e.g., in studies of
diffusive shock acceleration (DSA), in the form of a power-law dependence
of particle dispersion $\sqrt{\left\langle x^{2}\right\rangle }\propto t^{\alpha}$,
with $1/2<\alpha<1$ (superdiffusion) or $0<\alpha<1/2$ (subdiffusion).
The exact FP propagation leads to $\sqrt{\left\langle x^{2}\right\rangle }$
that smoothly changes from the ballistic ($\alpha\to1$) to diffusive
($\alpha\to1/2$) propagation with no dwelling at any particular value
of $\alpha$ between these limits. However, certain types of scattering
fields in shock environments, e.g., \cite{MD06}, may result in both
superdiffusive (L\'evy flights) and subdiffusive (long rests) transport
anomalies. Such fields are, however, less generic than those leading
to an isotropic scattering considered in this paper. They should perhaps
be justified on a case-by-case basis.

\section*{Acknowledgements}

\selectlanguage{british}%
This work was supported by the NASA Astrophysics Theory Program under
Grant No. NNX14AH36G.
\selectlanguage{english}%

%\section*{References}

\bibliographystyle{prsty.bst}
\bibliography{FPsolArxiv.bbl}

\end{document}